\documentclass{ieeeaccess}
\usepackage{cite}
\usepackage{amsmath,amssymb,amsfonts}
\usepackage{algorithmic}
\usepackage{graphicx}
\usepackage{textcomp}
\def\BibTeX{{\rm B\kern-.05em{\sc i\kern-.025em b}\kern-.08em
    T\kern-.1667em\lower.7ex\hbox{E}\kern-.125emX}}

\usepackage[printonlyused]{acronym}
\usepackage[T1]{fontenc}
\usepackage[utf8]{inputenc}
\usepackage{soul,xcolor}
\usepackage{url}
\usepackage{subcaption}
\usepackage{comment}
\usepackage{comment}
\usepackage{float}
\usepackage{caption}

\usepackage{tikz}
\NewSpotColorSpace{PANTONE}
\AddSpotColor{PANTONE} {PANTONE3015C} {PANTONE\SpotSpace 3015\SpotSpace C} {1 0.3 0 0.2}
\SetPageColorSpace{PANTONE}%
\usepackage{adjustbox}
\usetikzlibrary{matrix,positioning,arrows,shapes.geometric,arrows.meta,decorations.pathreplacing,quotes,shapes.misc}
\usetikzlibrary{decorations.markings}
\usepackage{pst-func, pst-plot}
\tikzset{
	woman/.pic={    
		\node[transform shape, circle,fill,minimum size=4.5mm] (head) at (0,0) {};
		\node[transform shape, draw, fill, trapezium, trapezium angle=55, trapezium stretches=true, rounded corners=2pt, minimum width=0.7cm, minimum height=1cm,
		,below = 1pt of head, inner sep=1pt] (body) {};
		\draw[transform shape, line width=1.5mm,round cap-round cap] ([shift={(-1mm,1mm)}]body.south) --++(-90:6mm);
		\draw[line width=1.5mm,round cap-round cap] ([shift={(1mm,1mm)}]body.south) --++(-90:6mm);
		\draw[line width=1.25mm,round cap-round cap, rounded corners] ([yshift=-.5\pgflinewidth]body.north) --++(2.5mm,0)--++(-75:6mm);
		\draw[line width=1.25mm,round cap-round cap, rounded corners] ([yshift=-.5\pgflinewidth]body.north) --++(-2.5mm,0)--++(255:6mm);
}}
\tikzset{
	man/.pic={    
		\node[transform shape, circle,fill,minimum size=4.5mm] (head) at (0,0) {};
		\node[transform shape, draw, fill, trapezium, trapezium angle=0, trapezium stretches=true, rounded corners=2pt, minimum width=0.4cm, minimum height=0.7cm,
		,below = 1pt of head, inner sep=1pt] (body) {};
		\draw[transform shape, line width=1.5mm,round cap-round cap] ([shift={(-1mm,1mm)}]body.south) --++(-90:9mm);
		\draw[line width=1.5mm,round cap-round cap] ([shift={(1mm,1mm)}]body.south) --++(-90:9mm);
		\draw[line width=1.25mm,round cap-round cap, rounded corners] ([yshift=-.5\pgflinewidth]body.north) --++(2.5mm,0)--++(-75:6mm);
		\draw[line width=1.25mm,round cap-round cap, rounded corners] ([yshift=-.5\pgflinewidth]body.north) --++(-2.5mm,0)--++(255:6mm);
}}
\tikzset{terminal/.style = {draw, shape = circle , thick, radius = 2cm},
	area/.style = {draw, shape = regular polygon, regular polygon sides = 6, thick, minimum width = 10cm},
	basestation/.style = {draw, shape = dart,fill=white, shape border rotate = 90, thick, minimum width = 0.5cm, minimum height = 0.5cm},
	transmission/.style = {decorate, decoration = {expanding waves, angle = 7, segment length = 4}, thick},
	label/.style = {font=\footnotesize}
}
\usepackage[flushleft]{threeparttable} 

\ifCLASSINFOpdf
\graphicspath{{pdf/}{jpeg/}{eps/}}
\DeclareGraphicsExtensions{.pdf,.jpeg,.eps}
\else

\DeclareGraphicsExtensions{.pdf,.jpeg,.png,.eps}
\else
\usepackage[dvips]{graphicx}
\fi

\begin{document}
\history{Date of publication xxxx 00, 0000, date of current version xxxx 00, 0000.}
\doi{10.1109/ACCESS.2019.DOI}

\title{A Machine Learning Framework for Sleeping Cell Detection in a Smart-city IoT Telecommunications Infrastructure}
\author{
\uppercase{Orestes~Manzanilla-Salazar}\authorrefmark{1}, \IEEEmembership{Member, IEEE},
\uppercase{Filippo Malandra}\authorrefmark{2}, \IEEEmembership{Member, IEEE},
\uppercase{Hakim Mellah}\authorrefmark{3}, 
\uppercase{Constant Wett\'e}\authorrefmark{4},
\uppercase{Brunilde~Sans\`o}\authorrefmark{5}, \IEEEmembership{Senior Member, IEEE}
}

\address[1]{Polytechnique Montreal, Montreal, QC H3T 1J4 Canada (e-mail: orestes.manzanilla@polymtl.ca)}
\address[2]{Department of Electrical Engineering, University at Buffalo, Buffalo, NY 14260 - 1920 USA (e-mail: filippom@buffalo.edu)}
\address[3]{Polytechnique Montreal, Montreal, QC H3T 1J4 Canada (e-mail: hakim.mellah@polymtl.ca)}
\address[4]{Ericsson, Stockholm, 164 40 Sweden (e-mail: constant.wette.tchouati@ericsson.com)}
\address[5]{Polytechnique Montreal, Montreal, QC H3T 1J4 Canada (e-mail: brunilde.sanso@polymtl.ca@polymtl.ca)}


\markboth
{Manzanilla-Salazar \headeretal: A Machine Learning Framework for Sleeping Cell Detection in a Smart-city IoT Telecommunications Infrastructure}
{Manzanilla-Salazar \headeretal: A Machine Learning Framework for Sleeping Cell Detection in a Smart-city IoT Telecommunications Infrastructure}

\corresp{Corresponding author: Orestes~Manzanilla-Salazar (e-mail: orestes.manzanilla@polymtl.ca).}

\begin{abstract}
The smooth operation of largely deployed  \ac{IoT} applications will depend on, among other things, effective  infrastructure failure detection. Access failures in wireless network \acp{BS} produce a phenomenon called ``sleeping cells'', which can render a cell catatonic without triggering any alarms or provoking immediate effects on cell performance, making them difficult to discover. To detect this kind of failure, we propose a \ac{ML} framework based on the use of \acp{KPI} statistics from the \ac{BS} under study, as well as those of the neighboring \acp{BS} with \emph{propensity} to have their performance affected by the failure. A simple way to define neighbors is to use adjacency in Voronoi diagrams. In this paper, we propose a much more realistic approach based on the nature of radio-propagation and the way devices choose the \ac{BS} to which they send access requests. We gather data from large-scale simulators that use real location data for \acp{BS} and \ac{IoT} devices and pose the detection problem as a supervised binary classification problem. We measure the effects on the detection performance by the size of time aggregations of the data, the level of traffic and the parameters of the neighborhood definition. The Extra Trees and Naive Bayes classifiers achieve \ac{ROC} \ac{AUC} scores of $0.996$ and $0.993$, respectively, with \acp{FPR} under $5 \%$. The proposed framework holds potential for other pattern recognition tasks in smart-city wireless infrastructures, that would enable the monitoring, prediction and improvement of the \ac{QoS} experienced by \ac{IoT} applications.
\end{abstract}


\titlepgskip=-15pt

\maketitle

\acresetall

\section{Introduction}\label{section:intro}

\PARstart{T}he deployment of the \ac{IoT} in urban areas is enabling  the creation of so-called ``smart cities'' where city life will be improved by using large amounts of information coming from hundreds of thousands of geographically distributed communicating devices. 
This information will lead to the automation of some systems  and the creation of new applications that will enhance city living.
Smart parking, smart pedestrian crossings, intelligent transportation systems, and intelligent power distribution are just a few of the new types of innovations that can be put in place with the effective exchange of information between city \ac{IoT} devices. \ac{IoT}-enabled data and services in smart cities rely on either (a) users interacting with smart devices connected to the Internet or (b) users using network services that depend on \ac{IoT} devices serving as sensors or actuators \cite{jin14}. In both cases, communications are essential for the \ac{IoT} applications to work.

Even though several telecommunication technologies have been proposed for the deployment of different IoT applications in cities \cite{zanella14,lin17}, the ubiquity of cellular communications is making operators and standardization entities such as the \ac{3GPP} push for a common cellular  infrastructure for smart cities based on \ac{4G} enhancements and  \ac{5G}. 

Even with the use of a common communication infrastructure, there are several  drawbacks of smart-city large-scale implementation. 
First, it heavily depends on reliable telecommunications, as even banal failures may lead to the massive malfunctioning of key automated systems. 
Second, the type of telecommunication traffic produced in smart cities will mostly be produced by \ac{IoT} machines inside those automated systems. 
The problem is that the statistical behaviour of this traffic is quite different from that produced by humans \cite{malandra18}, and the lack of direct human interaction will make it even more difficult to detect telecommunication failures. 
Finally, the distributed nature of the applications and the large number of devices and connections will also hinder failure detection.

One of the most difficult types of failure to detect in cellular networks is the so-called ``sleeping cell'' failure. 
It consists of failures that will not set-off alarms even if the cell is malfunctioning.
In human cellular communications, a sleeping cell will cause users to react to the lack of service, change location and eventually notify the operator.
This failure can be prolonged, in some cases days, before being detected by the operator, and corrective measures are taken\cite{barco12}. 
The influence of sleeping cell failures is greatly amplified in smart cities, where many automated systems may depend on the normal function of a particular cell.  
Thus, the city does not have the luxury of waiting several days for the malfunctioning cell  to be detected. 
The delay constraints of essential smart-city applications might be difficult to satisfy even with fully-operational \acp{BS} due to the massive number of devices that are expected to request access \cite{polese16}.

The objective of this paper is to present
a \ac{ML} framework to detect sleeping cells in a smart-city \ac{IoT} context. 
The framework is based on the following:

\begin{itemize}
\item the introduction of a novel concept of neighborhood between \acp{BS}, and
\item the use of aggregated \acp{KPI} over time intervals for different types of \ac{IoT} applications.
\end{itemize}

The data used to feed our framework were extracted from a large-scale \ac{IoT} infrastructure simulator that takes as its input a real city database of geographical locations of potential \ac{IoT} devices and the current locations and features of the \acp{BS} of several service providers.

In the remainder of this paper, we present the state of the art in Section~\ref{section:literature}. In Section~\ref{section:system}, we present the modeling of the system, emphasizing the relationship between the infrastructure technology and the locations of the \ac{IoT} devices. 
The \ac{ML} framework is detailed in Section~\ref{section:methodology}, starting with novel definitions of the cell neighborhood and proximity  that are at the core of the framework, followed by the simulation and \ac{ML} methodologies, and ending with some remarks on the implementation.
Numerical results are shown and commented on Section~\ref{section:results}, and conclusions are presented in Section~\ref{section:conclusion}.

\section{State of the art}\label{section:literature}

%

Failure detection of network elements is one of the main concerns of mobile network operators.
Several papers in the literature address this problem using real network operator data at the \ac{BS} level  \cite{coluccia13,shafiq12,shafiq13,delabandera15,rezaei16}.
This approach produces very accurate results for the specific networks, but the solutions are not easily generalized  due to the difficulty in retrieving real cellular network data. As a consequence, most authors use simulated data, such as in  \cite{khanafer08, mueller08,chernogorov11, 
panda12, 
ma13,  coluccia13, delabandera15, gomez16b, sun17,  manzanilla19}, though emulations based on real data can also be found \cite{khatib15}. In this work, we employ simulated network data generated with a large-scale network simulator \cite{sanso19} (an extension of  \cite{malandra18}), which employs real data on the positions of network elements and the parameters of the communicating nodes.


There has been much work dedicated to solving problems in wireless networks using \ac{ML}; for an extensive survey, see  \cite{zhang19b}. A common approach in failure detection is to \emph{learn} standard traffic patterns and quickly detect deviations from normal behavior (i.e., unsupervised learning). In particular, existing research focuses on i) anomaly detection \cite{cheung06,liao12,coluccia13}, ii) \acp{KPI} \cite{chernogorov16}, iii) clustering \cite{rezaei16, ma13,chernogorov11},  and  iv) dimensionality-reduction techniques \cite{chernogorov11, gomez16}. 
Other authors exploit known properties of cellular networks to perform supervised learning to detect faulty elements in a network \cite{liao12,rezaei16,liu18,sun17,mueller08,manzanilla19}. 

A complementary approach to ML proposed by \cite{khatib17b} is  to acquire data from troubleshooting (human) experts in mobile networks and to use their experience and knowledge to improve fault detection. In addition to the proposed techniques, fuzzy models can be used for failure detection, as presented in \cite{khatib15,gomez16b}.

Finally, some authors propose detecting failures in a network element by looking at anomalies in the traffic and \acp{KPI} from neighboring cells \cite{delabandera15,mueller08,manzanilla19}. This is particularly powerful when the traffic generated in a defected cell does not present remarkable anomalies in its \acp{KPI}, such as in the case of \ac{RACH}-sleeping cells, where new users cannot connect but existing users in the cell can continue to transmit regularly during a failure.

%
%
%
%

In this paper, we propose using well-known supervised learning techniques for \ac{BS} failure detection in a smart-city cellular infrastructure. In particular, for each cell, \acp{KPI} from neighboring cells are analyzed to highlight anomalies and detect defective \acp{BS}. Different from the reviewed literature, i) we consider advanced propagation models based not only on distance but also on other parameters, such as \ac{RSS}, the bandwidth, frequency, and antenna orientation, and ii) we define different neighbor categories to improve failure detection.


%
\section{System modelling}\label{section:system}

Let us first mention that we provide in Table \ref{tab:notation} a summary of the mathematical notation used in our modelling of the system as well as in the description of the proposed framework.

\subsection{Communication infrastructure}\label{subsubsection:communication}

The cellular network model is composed of a set $\mathcal{B}$ of base stations enumerated as $\lbrace b_1, \dots, b_M \rbrace$, a set of \ac{IoT} devices in geographical locations $\lbrace g_1, \dots, g_L \rbrace \subset \mathcal{G}$ , a backbone $\mathcal{N}$, and a \ac{DMC}. Only the uplink performance is considered; the core and metropolitan part of the network are modeled as a black box. 


\begin{table}[]
\caption{Summary of mathematical notation.}
\begin{tabular}{c|l}
\hline
\textbf{Symbol}     & \multicolumn{1}{c}{\textbf{Description}}                                                                                                                                                                                   \\ \hline
$\mathcal{B}$       & Set of $M$ \acp{BS} $\{b_1,...,b_M\}$.                                                                                                                                                                   \\ \hline
$\mathcal{G}$       & \begin{tabular}[c]{@{}l@{}}Set of potential geographical locations for an \ac{IoT} device.\end{tabular}                                                                         \\ \hline
$r_g(i)$            & \begin{tabular}[c]{@{}l@{}}\ac{RSS} from a $b_i \in \mathcal{B}$  as measured from location $g \in \mathcal{G}$\end{tabular}                                                                                     \\ \hline
$\mathbf{s}_g$      & \begin{tabular}[c]{@{}l@{}}Ordered list of size $\xi$ of the $b_{g,i} \in \mathcal{B}$ with \\highest $r_g(g,i)$ values in decreasing order\\$\{b_{g,1},...,b_{g,\xi}\}$. Subscript $i$ indicates priority.\end{tabular} \\ \hline
$f(i)$              & Failure indicator for $b_i \in \mathcal{B}$.                                                                                                                                                                               \\ \hline
$Pr(X = x)$         & Probability of event $x$.                                                                                                                                                                                                  \\ \hline
$\mathcal{C}^n_i$   & \begin{tabular}[c]{@{}l@{}}Set of $b_j \in \mathcal{B}$ belonging to the \\ \emph{Neighborhood Category} $n$ of $b_i \in \mathcal{B}$.\end{tabular}                                                    \\ \hline
$\mathcal{I}_{u-v}$ & \begin{tabular}[c]{@{}l@{}}Ordered list of consecutive positions\\ $\{u,u+1,...,v\}$, such that $\xi \geq v \geq u \geq 1$.\end{tabular}                                                                                   \\ \hline
$c^n_{i,j}$         &\begin{tabular}[c]{@{}l@{}} Category $n$ neighborhood indicator\\for $b_i, b_j \in \mathcal{B}$.\end{tabular}                                                                                                                                                                                                                                          \\ \hline
$P^{u-v}(b_i,b_j)$  & \begin{tabular}[c]{@{}l@{}}Indicator of $u-v$ \emph{Proximity} for a pair of\\ antennas $b_i, b_j \in \emph{B}$.\end{tabular}                                                                         \\ \hline
$x_{i,t}$           & Feature vector for $b_i \in \mathcal{B}$ in interval $t$.                                                                                                                                                                  \\ \hline
$o(x_{i,t})$        & \begin{tabular}[c]{@{}l@{}}Target value used for training of \ac{ML} models.\end{tabular}                                                                                                               \\ \hline
$\mathcal{N}$       & Backbone of the cellular network.                                                                                                                                                                                          \\ \hline
\end{tabular}
\label{tab:notation}
\end{table}

We assume a limited number of wireless channels (i.e., the \acp{RB} in \ac{LTE}) can be used to transmit data between users and \acp{BS}. 
This is done through dedicated control channels allocated through a random access procedure (\ac{RACH}), based on preamble transmissions. The available preambles are limited and might collide, triggering retransmission and introducing additional delay in the communications between devices and \acp{BS}.

The key parameters in this study are the collision probability and the access delay, i.e., the time required for a user packet to be received by the associated \ac{BS}. In particular, high-order statistics on those two parameters are used to detect sleeping cells. Further details on the methodology are provided in Section \ref{section:methodology}.

\subsection{Topology definition}\label{subsection:topology}

\begin{figure}
	\centering
	\begin{tikzpicture}
	\draw[draw=black] (-2,-2.5) rectangle ++(7,4.5);
	\node[inner sep=0pt] (tower2_1) at (2,1.5)
	{\includegraphics[width=.03\textwidth]{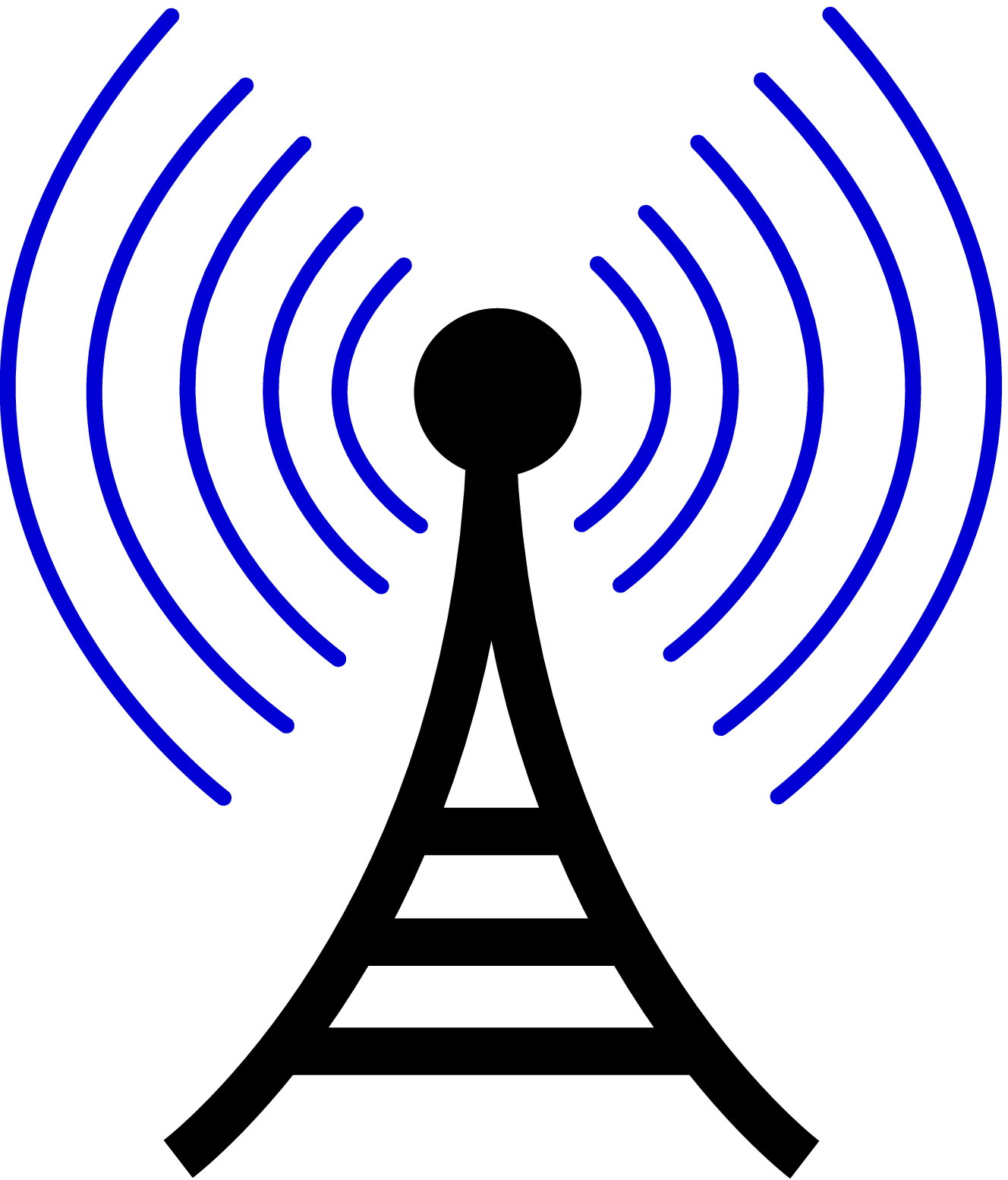}};
	\node[red,below = 2pt of tower2_1,scale=0.4]  {\Large\textbf{BS $i$}};
	\node[inner sep=0pt] (tower2_2) at (2,-1.5)
	{\includegraphics[width=.03\textwidth]{Wireless_tower.eps}};
	\node[red,below = 2pt of tower2_2,scale=0.4]  {\Large\textbf{BS $j$}};
	\node[inner sep=0pt] (tower2_3) at (-1,0)
	{\includegraphics[width=.03\textwidth]{Wireless_tower.eps}};
	\node[red,below = 2pt of tower2_3,scale=0.4]  {\Large\textbf{BS $k$}};
		\draw[draw=black] (1,-.25) -- (5,-.25);
		\draw[draw=black] (-0.5,2) -- (1,-.25);
		\draw[draw=black] (-0.5,-2.5) -- (1,-.25);
	
		\node[inner sep=0pt] (m11) at (-1,1.5)
		{\includegraphics[width=.03\textwidth]{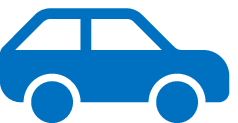}};
		\node[inner sep=0pt] (m12) at (-1,-2)
		{\includegraphics[width=.03\textwidth]{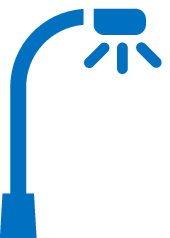}};
		\node[inner sep=0pt] (m13) at (-.5,1)
		{\includegraphics[width=.03\textwidth]{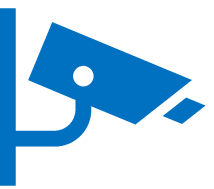}};
		\node[inner sep=0pt] (m14) at (0.1,.5)
		{\includegraphics[width=.03\textwidth]{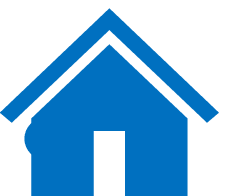}};
		\node[inner sep=0pt] (m15) at (0.1,-.5)
		{\includegraphics[width=.03\textwidth]{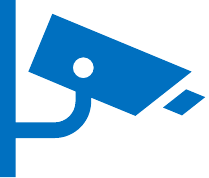}};
		\node[inner sep=0pt] (m16) at (-1.5,.8)
		{\includegraphics[width=.03\textwidth]{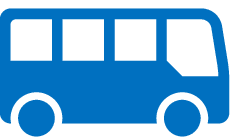}};
		\draw[blue] (m11) -- (tower2_3);
		\draw[blue] (m12) -- (tower2_3);
		\draw[blue] (m13) -- (tower2_3);
		\draw[blue] (m14) -- (tower2_3);
		\draw[blue] (m15) -- (tower2_3);
		\draw[blue] (m16) -- (tower2_3);
		
		\node[inner sep=0pt] (m21) at (0.5,1.5)
		{\includegraphics[width=.03\textwidth]{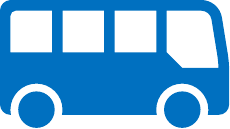}};
		\node[inner sep=0pt] (m22) at (0.6,1)
		{\includegraphics[width=.03\textwidth]{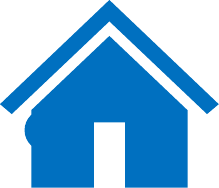}};
		\node[inner sep=0pt] (m23) at (1.3,0.5)
		{\includegraphics[width=.03\textwidth]{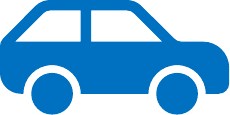}};
		\node[inner sep=0pt] (m24) at (3,0.6)
		{\includegraphics[width=.03\textwidth]{securitycamera}};
		\node[inner sep=0pt] (m25) at (4,.2)
		{\includegraphics[width=.03\textwidth]{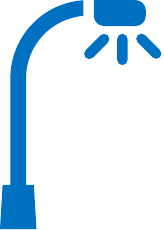}};
		\node[inner sep=0pt] (m26) at (4,1.5)
		{\includegraphics[width=.03\textwidth]{houseicon}};
		\draw[blue] (m21) -- (tower2_1);
		\draw[blue] (m22) -- (tower2_1);
		\draw[blue] (m23) -- (tower2_1);
		\draw[blue] (m24) -- (tower2_1);
		\draw[blue] (m25) -- (tower2_1);
		\draw[blue] (m26) -- (tower2_1);
		
		\node[inner sep=0pt] (m31) at (0.7,-1.8)
		{\includegraphics[width=.03\textwidth]{houseicon}};
		\node[inner sep=0pt] (m32) at (0.9,-1.2)
		{\includegraphics[width=.03\textwidth]{caricon}};
		\node[inner sep=0pt] (m33) at (1.5,-0.5)
		{\includegraphics[width=.03\textwidth]{bus}};
		\node[inner sep=0pt] (m34) at (3,-0.6)
		{\includegraphics[width=.03\textwidth]{houseicon}};
		\node[inner sep=0pt] (m35) at (4,-.7)
		{\includegraphics[width=.03\textwidth]{securitycamera}};
		\node[inner sep=0pt] (m36) at (4,-1.5)
		{\includegraphics[width=.03\textwidth]{trafficlight}};
		\draw[blue] (m31) -- (tower2_2);
		\draw[blue] (m32) -- (tower2_2);
		\draw[blue] (m33) -- (tower2_2);
		\draw[blue] (m34) -- (tower2_2);
		\draw[blue] (m35) -- (tower2_2);
		\draw[blue] (m36) -- (tower2_2);
	\end{tikzpicture}
	\caption{A sample scheme of the proposed architecture with three \acp{BS} and a large number of \ac{IoT} devices.}
	\label{figure:comm_scenario}
\end{figure}

%
The framework was built with real telecommunications and urban data from the city of Montreal (see details in \cite{malandra18}).
In Fig. \ref{figure:comm_scenario}, a toy example of a smart-city cellular system is displayed: network users are represented by \ac{IoT} devices, such as cars, buses, traffic lights, and security cameras. 
Details on the types of \ac{IoT} device considered and their characteristics can be found in a previous work \cite{malandra17b}, where six different \ac{IoT} applications are presented. The rectangle represents the geographical boundaries of a smart city, in which three \acp{BS} $b_i$, $b_j$, and $b_k$ are installed and provide network access to the \ac{IoT} devices. 
The geographical position and other features of the \acp{BS}, such as the bandwidth, transmitted power, and orientation, were retrieved from \cite{SMSD19}.

 To characterize the links between users and \acp{BS} we i) define a threshold on the received power, ii) compute the power received from each of the \acp{BS} by each \ac{IoT} device, and iii) determine the list of \acp{BS} that cover each \ac{IoT} device. A threshold of $-100$ dBm is considered in this study. The received power is computed according to the Cost-Hata (Cost-231 defined in \cite{akhpashev16}) propagation model, which allows computing the path loss based on key parameters, such as the frequency, distance, and height. This propagation model is also combined with the corresponding radiation patterns for each \ac{BS}. This leads to computing the \ac{EIRP}, based on the elevation, gain and inclination of the antennas, which are also available at \cite{SMSD19}.
The list of \acp{BS} covering a certain \ac{IoT} device can be very large, especially in a densely populated urban scenario like Montreal, and this can lead to computational inefficiencies and large execution times. As a consequence, this list is limited to the $\xi$ \acp{BS} with the highest received power. The list is used, as described in Section \ref{subsection:closeness}, to combine the analysis of one cell's \acp{KPI} with those of neighbor cells, and ultimately to detect the sleeping cells with high accuracy.

\subsection{The sleeping cell problem}\label{subsection:sleepingcell}

A sleeping cell is usually defined as a cell that is not entirely operational and whose malfunctioning is not easily detectable by the network operator, as highlighted in \cite{chernogorov16}. 
This term is generally used to describe a wide variety of hardware and software failures, which degrade the \ac{QoS} and \ac{QoE} and can remain hidden to the network operator for a long time (days or even weeks) \cite{hamalainen12}. 
In this study, we address a particular type of sleeping cells that affects the \ac{RACH} in \ac{LTE} networks \cite{chernogorov16}. 
On the one hand, this type of problem affects new users who are not able to complete the access procedure and consequently cannot access the network. 
On the other hand, existing users, which were already connected to the \ac{BS} when the problem manifested, continue to transmit. 
As a consequence, standard methods based on traffic monitoring fail to detect the problem, because the network operator continues to monitor updated statistics coming from the \textit{RACH-sleeping} \ac{BS}. 
Progressively, all the ongoing connections end, and the cell ceases all activity.

\section{A framework for sleeping cell detection}\label{section:methodology}

\subsection{Neighborhood/closeness definition}\label{subsection:closeness}

When trying to detect if a particular BS has failed, our key idea is to include data from its ``neighborhood''. However \emph{how does one determine which \acp{BS} can be considered as ``neighbors''?} 
Though the \ac{BS} distance can be used, as done in \cite{manzanilla19}, we now propose a richer definition: a neighboring \ac{BS} is actually one whose performance \acp{KPI} are \textit{likely} to be affected by the access failure in the  \ac{BS} under study. 
Accordingly, we base our definition on the following: \\

\begin{figure}[!t] 
	\centering
	\includegraphics[width=2.4in]{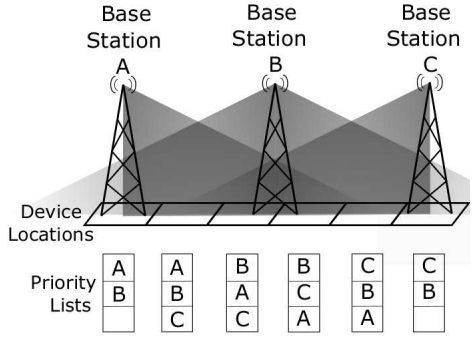}
	\centering	
	\caption{Priority list example.}
	\label{figure:priority_list}
\end{figure}

\noindent \textit{1. Antenna priority and \ac{RSS}}: when sending access requests, an \ac{IoT} device at location $g \in \mathcal{G}$ will choose the \ac{BS} $b_m \in \mathcal{B}$ such that:

\begin{equation}
\label{equation:max_rss}
	m = \underset{1\geq j \geq M}{\operatorname{argmax}} \{ r_g(j) \}
\end{equation}

where $r_g(j)$ is the signal strength of a \ac{BS} $b_j$ measured at location $g$.
The set of all the \acp{BS} considered as options for a device located at $g \in \mathcal{G}$ is depicted as a  \textit{priority list} $\mathbf{s}_g$ of size $\xi$, defined as the following sorted list of \acp{BS}:

\begin{equation}
\label{equation:priority_list_def}
	\mathbf{s}_g = [b_{g,1}, b_{g,2},..., b_{g,\xi}]
\end{equation}
where the \ac{RSS} values of the \acp{BS} $b_1 = b_{g,1}, b_2 = b_{g,2}, ..., b_{\xi} = b_{g,\xi}$, hold the following relationship:
\begin{equation}
	r_g(1) \geq r_g(2) \geq...\geq r_g(\xi)
\end{equation}
A device at location $g$ will send its access request to BS $b_{g,1}$ first. When the \ac{BS} fails, the next \ac{BS} on the list ($b_{g,2}$) is considered, and so on.   
Fig. \ref{figure:priority_list} shows examples of \acp{BS} and their positions in priority lists of size $\xi = 3$ for a series of locations. 
Note that in our implementation the $12$ antennas with the highest \ac{RSS} are considered ($\xi = 12$). 
In some locations, the priority list can be shorter, as a consequence of the threshold mentioned in Section \ref{subsection:topology}. 
The reader should be aware that even if \ac{RSS} decay is presented in Fig. \ref{figure:priority_list} as decreasing linearly with distance, this is done \emph{only to simplify} the explanation.\\
\\
	\noindent \textit{2. Directional antennas}: antenna tilt and orientation are important factors in the computation of the \ac{RSS} in the propagation model used in our simulator. Therefore, the strongest received signal might not come from the closest \ac{BS}. In Fig. \ref{figure:proximity2}, where a set of \acp{BS} numbered from $1$ to $17$ are shown. We observe that even though BS $6$ is closer than BS $15$ to the BS under study, it might not be the second BS in the priority list for any location, while BS $15$ effectively is. In our modeling of the system, this might happen because no location receives the signals of BS $6$ and the one under study as the two strongest ones, due to the antenna directionality and power values.\\
\\
	\noindent \emph{3. \acp{KPI} availability}: it is feasible to obtain aggregations of \acp{KPI} for all packets processed by a \ac{BS} during any period of time.

The type of failure we analyze in this paper is that of the sleeping cells, which are \acp{BS} whose \emph{access function becomes inactive}, affecting the performance of ``close'' \acp{BS}. This process can be described as follows:

\begin{itemize}

	\item \emph{Step 1}: The observed \ac{BS}'s access function fails. Ongoing transmissions continue to be served by the \ac{BS}.
	\item \emph{Step 2}: Idle devices that usually would request access to the failed \ac{BS}, choose as an alternative the one with the second-highest \ac{RSS}.
	\item \emph{Step 3}: The additional traffic, produced by the ``new'' devices requesting access, induces a performance degradation in the \emph{chosen} \ac{BS}.

\end{itemize}

Because of its degradation, it is of interest to include the \ac{BS} chosen in step 2 as a neighbor in the pattern analysis process. This is why we base our definition of \emph{neighborhood} on the notion of the \emph{probability of experiencing a performance degradation}. 
Note also that, even though simultaneous failures of nearby \acp{BS} may not be frequent, they cannot be ruled out. Therefore devices around the failed location may go down their priority list until they find an operational \ac{BS}. The farther down a \ac{BS} is in the priority list, the less likely  it is to receive the extra traffic, as it would require the \emph{simultaneous} failure of all the \acp{BS} positioned ``above'' it in the priority list.

\subsubsection*{Using probabilities to define neighborhood categories}
\label{subsubsection:probabilities}


We define now a novel idea to determine whether two 
\acp{BS} are ``neighbors'' or not, based on a threshold on the probability of each of the \acp{BS} affecting the other's performance in case of a failure. The following example with $\xi=4$ illustrates the intuition behind this approach. Let  $p$ be the probability of access failures of  \emph{any} \ac{BS} during any given time interval of duration $T$. Let us also assume that failures in different \acp{BS} and time intervals are independent. Given the device location $g \in \mathcal{G}$, let $\mathbf{s}_g = [b_{g,1}, b_{g,2}, b_{g,3}, b_{g,4}]$ be the priority list  containing the 4 \acp{BS} with the highest received power at $g$. When  $b_{g,1}$ fails, one of the following occurs:

\begin{itemize}
	\item $b_{g,2}$ is operational with probability $(1-p)$ and it will receive all the traffic from devices at $g$ with probability $1(1-p)$.
	\item $b_{g,2}$ is asleep, which will happen with probability $p$, $b_{g,3}$ is operational with probability $(1-p)$, and the requests for access of the device at $g$ will be handled by $b_{g,3}$. This will occur with probability $p(1-p)$.
	\item both $b_{g,2}$ and $b_{g,3}$ are asleep with probability $p^2$, $b_{g,4}$ is operational with probability $(1-p)$, and $b_{g,4}$ will be the alternative \ac{BS} receiving the access requests. This will happen with probability $p^2(1-p)$.
\end{itemize}

This example shows the intuition behind our definition of \emph{neighborhood of category $n$ of \ac{BS} $b_k$}. A \ac{BS} is a \emph{category} $n$ neighbor of \ac{BS} $b_k$ when it is the $n$-th option to request access if \ac{BS} $b_k$ fails.

To formalize the relationship between the probability of receiving traffic normally served by $b_{g,1} \in \mathbf{s}_g$, and the condition of being a neighbor of category $n$, let us first define the failing state indicator of the \ac{RACH} function of \ac{BS} $b_i\in\mathcal{B}$ as:

\begin{equation}
\label{equation:failing_state}
	f(b_i) =
	\begin{cases}
		1 & \text{if access function of }b_i\text{ is inactive}\\
		0 & \text{otherwise}
	\end{cases}
\end{equation}
Given an ongoing access failure in $b_{g,1}  \in \mathbf{s}_g$, the probabilities of failure for the \acp{BS} in $\mathbf{s}_{g}$ are:
\begin{eqnarray}
	Pr(f(b_{g,1})=1)&=& 1  \\
	Pr(f(b_{g,n}=1)&=& p; \forall n=2,...,\xi
\end{eqnarray}

When the access function of \ac{BS} $b_i= b_{g,1}$ fails, we can define $Q(i,j)$ as the probability of \ac{BS} $b_j = b_{g,n}  \in \mathbf{s}_g$ receiving traffic normally served by \ac{BS} $b_i$. This probability can be modeled as:

\begin{align}
\label{equation:degradation_probability}
&Q(i,j) = Pr(f(b_{g,1} = 1) \cap ... \nonumber\\
&\; ...\cap Pr(f(b_{g,n-1}=1) \cap Pr(f(b_{g,n}=0)\\
&\:=p^{n-1}(1-p) \nonumber
\end{align}

We now define $\mathcal{C}^n(b_i)$, the \emph{neighborhood category} $n$ as:

\begin{equation}
\label{equation:neighborhood_definition}
\mathcal{C}^n(b_i) = \{b_{j \neq i} | Q(i,j) \geq p^{n-1}(1-p) \}
\end{equation}

As a consequence of Equation (\ref{equation:neighborhood_definition}), each neighborhood is nested inside those of higher  categories, following the structure of a set of  \emph{Russian dolls} (a.k.a. \textit{Matryoshkas} or \textit{Babushkas}):

\begin{equation}
\label{equation:matrioshkas}
\mathcal{C}^1(b_i) \subset ... \subset \mathcal{C}^{\xi-1}(b_i)
\end{equation}

Applying the definitions to the example, assuming that there is only one device location in the system $g \in \mathcal{G}$, we obtain the following possible neighborhood category sets  for $b_i=b_{g,1}$:

\begin{align}
\label{equation:neighborhood_examples}
&\mathcal{C}^1(b_{g,1}) = \{b_{g,2}\} \nonumber\\
&\mathcal{C}^2(b_{g,1}) = \{b_{g,2},b_{g,3}\}\\
&\mathcal{C}^3(b_{g,1}) = \{b_{g,2},b_{g,3},b_{g,4}\} \nonumber
\end{align}

	where $b_{g,1}, b_{g,2}, b_{g,3}, b_{g,4} \in \mathbf{s}_g \subset \mathcal{B}$.

\begin{figure}[!t] 
	\centering
	\includegraphics[width=2.5in]{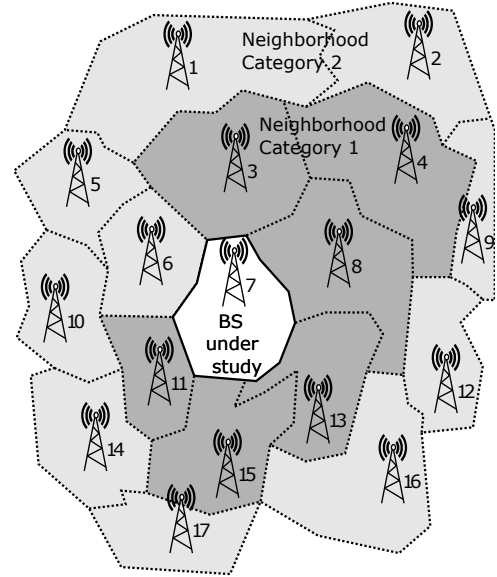}
	\centering	
	\caption{Example of neighborhood categories 1 and 2.}
	\label{figure:proximity2}
\end{figure}

Note that in Fig. \ref{figure:proximity2}, \ac{BS} 7 is the target of this analysis, and the \acp{BS} in neighborhood category 1 (dark gray) are part of the set of category 2 (light gray), as a consequence of Equation (\ref{equation:matrioshkas}). To highlight the differences between the proposed neighboring structure and the classical geographical one, in this example, an immediate neighbor, such as \ac{BS} $6$,  is excluded from the neighborhood of category 1, and the more distant \ac{BS} $4$ instead belongs to it. 
This choice is to emphasize the fact that distance is not the criterion to determine the membership of a neighborhood set, which is actually determined by the position in priority lists. 

\subsubsection*{The u-v proximity}
\label{subsubsection:proximity}

To find the set of neighbors for each target \ac{BS}, we need to first compute the $u-v$ \emph{proximity}. 
Two \acp{BS} \emph{have} $u-v$ ($u < v$) proximity \emph{if} there exists at least one device location such that in its priority list, the positions occupied by the two \acp{BS} are between the $u^{th}$ and $v^{th}$ positions (including the extremes). 

Based on this definition, multiple $u-v$ proximities can be defined for a single pair of \ac{BS} if there is more than one location whose list contains antennas from both \ac{BS}. 
This occurs because a pair of \acp{BS} can occupy very diverse positions in the priority lists  in different device locations.
At a location well positioned to receive signals from both \acp{BS}, both might occupy the first two positions of the list. At a location far from both \acp{BS}, they might occupy the two last positions of the list.

The existence of multiple $u-v$ proximities for the same pair of \acp{BS} is not necessarily a problem. Their usefulness becomes evident when we consider that the signals from a single pair of \acp{BS} might not be received with enough strength in any device to have $1-2$ proximity, for example, but are received strongly enough in at least one device to have $2-3$ proximity. This allows us to say that these \acp{BS} do not belong to each other's neighborhood category 1 but that they belong to each other's neighborhood category 2.  No device normally connected to one of them will have as a first choice the other \ac{BS} in case of an access failure, unless there are two simultaneous failures.

\begin{figure}[!t] 
	\centering
	\includegraphics[width=2.7in]{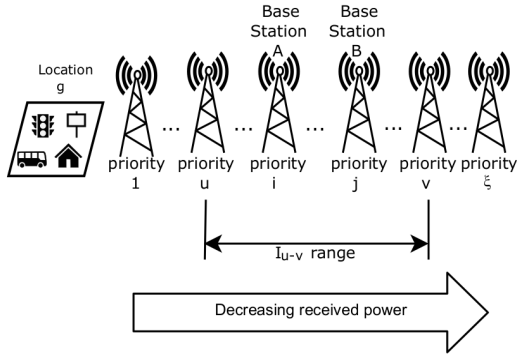}
	\centering	
	\caption{Range u-v of proximity between \acp{BS} (i,j).}
	\label{figure:uv_closeness}
\end{figure}

We can formally define the $u-v$ \emph{range} of priorities as the following ordered set:

\begin{equation}
\label{equation:u_v_range}
	\mathcal{I}_{u-v} = \{ u, u+1,...,v \}
\end{equation}	
where $\xi \geq v > u \geq 1$.

%

Let the $u-v$ \emph{proximity indicator} of a pair of \acp{BS} $b_i, b_{j} \in \mathcal{B}$ be defined as follows:

\begin{equation}
\label{equation:u_v_prox_ind}
	P^{u-v}(b_i,b_j) =
		\begin{cases}
			1 & \text{if } \exists g \in \mathcal{G}:\,\, b_i,b_j \in \{b_{g,u},...,b_{g,v}\} \\
			0 & \text{otherwise}
		\end{cases}
\end{equation}
where
\begin{description}
	\item $i \neq j$;
	\item $u,u+1,...,v \in \mathcal{I}_{u-v}$;
	\item $b_i, b_j,   \in \mathcal{B}$; and
	\item $b_{g,u},...,b_{g,v} \in \mathbf{s}_g \subset \mathcal{B}$.

\end{description}

In Fig. \ref{figure:uv_closeness}, to visually show the $u-v$ proximity concept, the following are displayed: location $g \in \mathcal{G}$; the set of machines installed in $g$; and some of the $\xi$ \acp{BS} in the priority list $\mathbf{s}_g$. Note that, in this example, $u$, $i$, $j$, and $v$ all belong to the range $\mathcal{I}_{u-v}$. Therefore, applying the definition in (\ref{equation:u_v_prox_ind}), we have that $P^{u-v}(A,B) = 1$, for \acp{BS} $A = b_{g,i}$ and $B = b_{g,j}$.

Note that the $u-v$ proximity is a symmetrical property, under the general assumption of the existence of at least one other location where the positions that both \acp{BS} occupy in a priority list are inverted. Therefore, for any pair of \acp{BS} $b_i,b_j \in \mathcal{B}$, we have that:

\begin{equation}
\label{equation:proximity_symmetry}
	P^{u-v}(b_i,b_j) =  P^{u-v}(b_j,b_i), \forall i\neq j
\end{equation}

The $u-v$ proximity is a property that can be used to determine whether any pair of \acp{BS} has \emph{high, medium or low propensity} of affecting each other's performance by considering a $u-v$ range that covers positions at the beginning, middle or last part of the priority list. In our specific implementation, we are interested in identifying those \acp{BS} that belong to a \emph{specific neighborhood category}. We can see in Equation (\ref{equation:neighborhood_definition}) that the \acp{BS} of interest for a particular $\mathcal{C}^{n}$ are receiving traffic from the target \ac{BS} with a probability of $p^{n-1}(1-p)$ \emph{or higher}. This means that the $u-v$ range for this application starts in the first position ($u=1$).

\begin{table}[]
\caption{Relation between priority, probability, category and proximity}
\begin{tabular}{c|c|c|c|c}
\hline
\textbf{Alternative}              & \textbf{Position}                         & \textbf{$Pr$}                     & \multicolumn{1}{c|}{\textbf{Neighborhood}} & \textbf{$u-v$}                        \\
\multicolumn{1}{c|}{\textbf{BSs}} & \multicolumn{1}{c|}{\textbf{in $\mathbf{s}_g$}} & \multicolumn{1}{l|}{\textbf{}} & \textbf{Categories}                        & \multicolumn{1}{l}{\textbf{interval}} \\ \hline
$b_{g,2}$ & 2 & \multicolumn{1}{l|}{$1(1-p)$}    & $\mathcal{C}^1$ & $1-2$ \\
$b_{g,3}$ & 3 &$p(1-p)$ & $\mathcal{C}^2$ & $1-3$                                   \\
$b_{g,4}$ & 4 & $p^{2}(1-p)$ & $\mathcal{C}^3$                                         & $1-4$                                  
\end{tabular}
\label{tab:relationships}
\end{table}

Table \ref{tab:relationships} illustrates where the $u-v$ range ends, in relationship to a particular neighborhood category. We can observe the relationship between:
\begin{enumerate}
	\item The \emph{position} of the \acp{BS} in a priority list,
	\item The probability they have of receiving access requests typically served by the \ac{BS} in the first position if it fails,
	\item The neighborhood of lowest category that includes each \ac{BS}, and
	\item the $u-v$ interval associated to each case.
\end{enumerate}
Table \ref{tab:relationships} was built following the toy example presented in section \ref{subsubsection:probabilities} to illustrate the intuition behind our approach, which was also used in Equation (\ref{equation:neighborhood_examples}).
Under the assumption that $b_{g,1}$ fails, each of the three \acp{BS} has decreasing probabilities of receiving access requests originally intended for the \ac{BS} in first position. We can appreciate as a general rule that \textit{for a neighborhood category} $n$:
\begin{itemize}
	\item \acp{BS} in positions $2,...,n+1$ are included in it.
	\item The lower bound of the probability these \acp{BS} have of receiving access requests from $b_{g,1}$ is $p^{n-1}(1-p)$.
	\item The associated proximity range ends in $n+1$.
\end{itemize}
We conclude that the $u-v$ range associated to a neighborhood category $n$ is $1-(n+1)$.

Formally:

\begin{equation}
\label{equation:neighborhood_proximity_relation}
	b_{j \neq i} \in \mathcal{C}^n(b_i) \Leftrightarrow  P^{1-(n+1)}(b_i,b_j) = 1
\end{equation}
for at least one location $g \in \mathcal{G}$. 

We can illustrate this relationship with the following example: if \ac{BS} $b_j$ belongs to the neighborhood  category $3$ of $b_i$ ($b_j \in \mathcal{C}^3(b_i)$), it means that $b_i$ and $b_j$ have  $1-4$ proximity ($P^{1-4}(b_i,b_j)=1$).

Because of the association defined between the notions of proximity and neighborhood, the symmetry defined in (\ref{equation:proximity_symmetry}) also implies a symmetry in neighborhood relationships such that:

\begin{equation}
\label{equation:neighborhood_symmetry}
	b_{j} \in \mathcal{C}^n(b_i) \Leftrightarrow  b_{i} \in \mathcal{C}^n(b_j) , \forall i\neq j
\end{equation}

\subsubsection*{Neighborhood matrices}
\label{subsubsection:neighborhood_matrices}

In the proposed framework, we aggregate \acp{KPI} of the ``neighbors'' of a \ac{BS} whose failing state we wish to study. To identify the neighbor \acp{BS}, we use a \emph{neighborhood indicator} for each pair of \acp{BS}. We arrange these indicators in an $M\times M$ \emph{neighborhood matrix}, where $M$ is the number of \acp{BS} in the system.

For a \emph{neighborhood category} $n$, each element of the $M\times M$ \emph{neighborhood matrix} $C^n$ is defined as:

\begin{equation}
\label{equation:neighborhood_matrix}
	c^n_{i,j}=
	\begin{cases}
		1& \text{if }b_j \in \mathcal{C}^n(b_i),j \neq i\\
		0& \text{otherwise} 
	\end{cases}
\end{equation}

Note that $c^n_{i,j} = 0$ when $i=j$ and that $c^n_{i,j} = c^n_{j,i}$ because of (\ref{equation:neighborhood_symmetry}).

\begin{figure*}[t!]
    \centering
    \begin{subfigure}[b]{0.5\textwidth}
        \centering
        \includegraphics[height=1.2in]{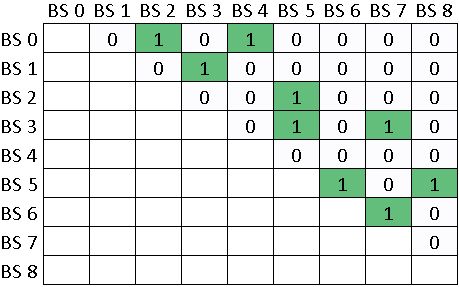}
        \caption{Neighborhood category 1.}
        \label{figure:neighborhood_1}
    \end{subfigure}%
    ~ 
    \begin{subfigure}[b]{0.5\textwidth}
        \centering
        \includegraphics[height=1.2in]{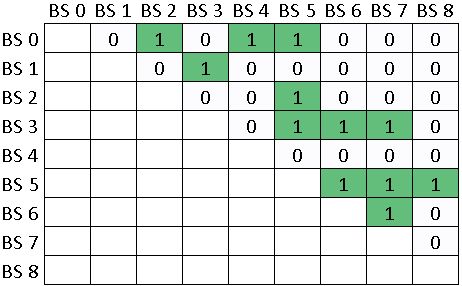}
        \caption{Neighborhood category 5.}
        \label{figure:neighborhood_5}
    \end{subfigure}
    \caption{Example of neighborhood matrices.}
\end{figure*}

The neighborhood matrices $C^1$ and $C^5$ computed in our implementation are partially shown in Figs. \ref{figure:neighborhood_1} and \ref{figure:neighborhood_5}. We can observe that the \ac{BS} identified with a $0$ is a neighbor of \acp{BS} $2$ and $4$ when considering a neighborhood of category $1$ (Fig. \ref{figure:neighborhood_1}). If we consider a neighborhood of category $5$, \ac{BS} $0$ is also a neighbor of \ac{BS} $5$. We can observe a similar situation for \acp{BS} $3$ and $5$, which have one more neighbor when the category is augmented to $5$. For the cellular network simulated, the matrices $C^n$ are of size $479 \times 479$, as there are $479$ \acp{BS} in the city.

%

The neighborhood matrices are computed once per  cellular infrastructure and recomputed only if there is a change in the \ac{RSS} values or if antennas or \acp{BS} are removed or added. The matrices are evaluated when aggregating the neighborhood \acp{KPI}  for each of the \acp{BS} as part of the construction of the ``feature vectors'' that allow the use of \ac{ML} for failure detection. As mentioned, a neighborhood is  associated with the probability of experiencing a performance degradation as a consequence of a failure, whose pattern we intend to detect. The specifics on how \acp{KPI} are aggregated can be found in Section \ref{subsection:ml_framework}.

\subsection{Network simulation}\label{subsection:simulation}

We use an \ac{LTE} simulator\footnotemark similar to the one described in \cite{malandra18}. The way \ac{IoT} devices gain access to the \acp{BS} is based on the computation of priority lists ($\xi = 12$) for each device served by the mobile infrastructure. The simulator allows for several propagation models to compute the \ac{RSS} values involved in the construction of the priority lists. Because this model encompasses the uplink \ac{RACH} procedure and the transmission until  reception at the \ac{EPC}, its output is composed of counters and statistics related to both phases:

\begin{itemize}
	\item Number of packets created.
	\item Number of packets transmitted.
	\item Number of \ac{RACH} collisions.
	\item Number of \ac{RACH} attempts.
	\item Minima, maxima and average of the \ac{RACH} delays.
	\item Minima, maxima and average of the transmission times (from \ac{RACH} completion to reception at the \ac{EPC}).
\end{itemize}

The priority lists are computed considering as options the antennas, instead of the \acp{BS}. In the preprocessing to compute the neighborhood matrices, the antennas identification codes in these lists is replaced by the identification of the \ac{BS} where the antennas are installed.

\footnotetext[1]{Smart cities M2M: \url{https://www.trafficm2modelling.com/}}

The \ac{RACH} failures are modeled as affecting simultaneously all the antennas of a particular \ac{BS}. Because \acp{BS} generally do not have the same number of antennas, most of the time, the probabilities for the positions in the priority lists may have values higher than the theoretical minimum  described in Section \ref{subsection:closeness} for a particular neighborhood category. We purposely choose to omit the implementation of countermeasures in the preprocessing to address the ``noise'' introduced by it, as the effect on the results does not hinder the methodology.

\subsubsection{Simulated scenarios}
\label{subsubsection:scenarios}

In Table \ref{tab:sim_scenarios}, we show the devices types, levels of traffic, duration of the simulation, total number of \acp{BS} and number of failing \acp{BS}.

\begin{table}[]
\centering
\caption{Simulation scenarios}
\begin{threeparttable}
\begin{tabular}{lc}
                                                                                              \hline
\multicolumn{1}{l|}{Type of device}                                                  &  \begin{tabular}[c]{@{}c@{}}smart meters, surveillance cameras, bus stops,\\ traffic lights, parking lots, microPMUs\tnote{*}\end{tabular} \\ \hline
\multicolumn{1}{l|}{Levels of traffic}                                                       &High / low                                                                                                                               \\ \hline
\multicolumn{1}{l|}{\begin{tabular}[c]{@{}l@{}}Simulation duration (h)\end{tabular}}                                                                                                                                                       & 12                                                                                                                                       \\ \hline
\multicolumn{1}{l|}{Total \acp{BS}}                                                           &479                                                                                                                                      \\ \hline
\multicolumn{1}{l|}{Failing \acp{BS}}                                                        & 50                                                                                                                                      
\end{tabular}
\begin{tablenotes}
	\item[*] Micro-phasor measurement units are power monitoring devices.
\end{tablenotes}
\end{threeparttable}
\label{tab:sim_scenarios}
\end{table}

We considered two scenarios in our simulations: high traffic and low traffic. In the high-traffic scenario, smart meters, parking slots, bus stops, surveillance cameras, and traffic lights generate packets at double rate with respect to the low-traffic scenario. Fire alarms and \acp{microPMU} generated packets at the same rate for both scenarios.

\subsubsection{RACH failure generation}

In a random sample of 50 \acp{BS}, total \ac{RACH} failures were parameterized to initiate at the beginning of each 1-hour simulation, with a duration of 30 minutes. This process was repeated 12 times (with different random seeds).
In every simulation, the first 10 minutes are used to initialize the network and their results are omitted from the KPIs aggregation.

\subsection{AN ML framework}\label{subsection:ml_framework}

\subsubsection{Preprocessing}\label{subsubsection:preprocessing}

The output of the simulator was aggregated in three ways: 
\renewcommand{\theenumi}{\roman{enumi}}%
\begin{enumerate}
	\item Across non-overlapping time-bins or intervals (according to the size of the time aggregations),
	\item across the antennas of each \ac{BS}, and 
	\item across the \ac{IoT} devices.
\end{enumerate}
As a result, the data set contained the statistics at the \ac{BS} level and considered generic traffic (without distinction among the traffic generated by the different devices/applications). The results were preprocessed aggregating the data at the \ac{BS} level in time intervals of 5, 10, 15 and 30 minutes. This allowed us to study the effect of aggregation size on detection performance.

A fundamental part of the preprocessing was the computation of the neighborhood matrices for each of the neighborhood categories. This process involves the following steps:

\begin{itemize}
	\item Analyzing the priority lists for each location in the network.
	\item Processing the priority lists to obtain the $u-v$ proximities between each pair of \acp{BS}.
	\item Using the $u-v$ proximities to obtain the neighborhood matrices.
\end{itemize}

Our aggregating procedure consisted of computing the following statistics: average, variance, skewness, kurtosis, percentile (5, 25, 50, 75, 95), minimum, maximum and range.

\begin{figure*}[h!] 
	\centering
	\includegraphics[width=6.8in]{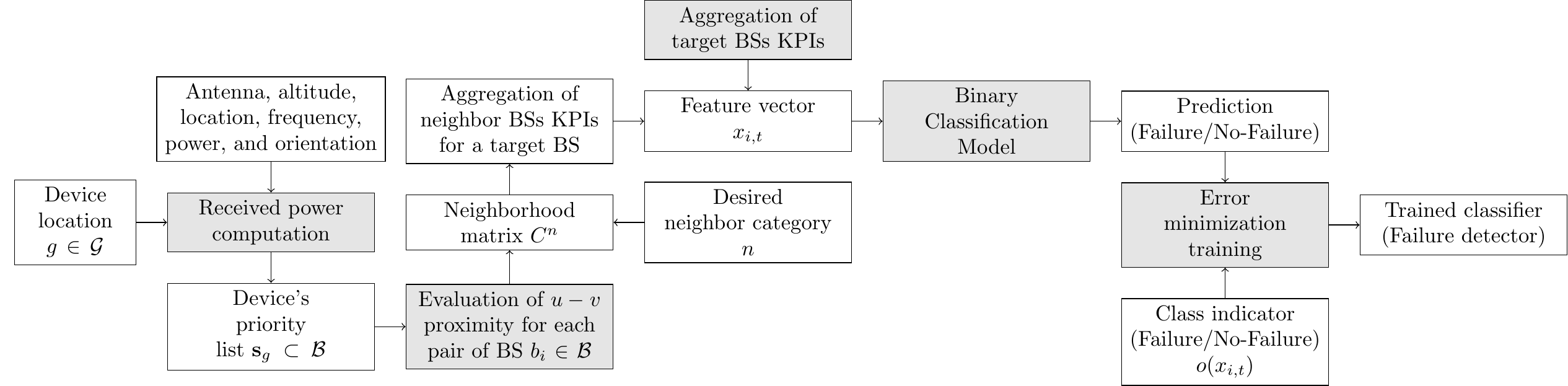}
	\centering	
	\caption{Graphical description of the proposed framework for failure detection.}
	\label{figure:process}
\end{figure*}

After the data were aggregated for each interval/\ac{BS}, a normalization step is used to force the values to lie within the range $(0, 1)$.

The feature vectors ${x_{i,t}}$ to be used in the \ac{ML} algorithms are built concatenating, for each aggregation interval $t$ and ``target'' \ac{BS} $b_i$ two vectors:

\begin{itemize}
	\item Aggregation of the statistics of \ac{BS} $b_i$ during interval $t$.
	\item Aggregation of the statistics of all the neighboring \acp{BS} in $\mathcal{C}^n(b_i)$ during interval $t$.
\end{itemize}

To perform supervised classification, the data set is completed by associating each feature vector $x_{i,t}$ to a category label, a class indicator or a ``target value'' $o(x_{i,t})$, defined as:

\begin{equation}
\label{equation:target_value}
	o(x_{i,t}) =
	\begin{cases}
		1 & \text{if }f(b_i) = 1\text{ at interval }t\\
		0 & \text{otherwise}
	\end{cases}
\end{equation}

This procedure is repeated using the data generated by the simulator for each time aggregation size and neighborhood category. The whole process, for a particular interval of time, is described in the diagram in Fig. \ref{figure:process}.

\subsubsection{Models and training strategy}\label{subsubsection:Training}

We report our results for  the following binary classifiers:

\begin{enumerate}
	\item Naive Bayes
	\item Logistic Regression
	\item Linear, Quadratic, Cubic and \ac{RBF} Support Vector Machines
	\item Decision Trees
	\item Extra Trees
	\item Bagged Decision Trees
	\item Random Forest
	\item Shallow (Single Hidden Layer) Neural Networks
\end{enumerate}

For each of the simulation scenarios and preprocessing strategies, the data set is randomly split into \emph{training} ($70 \%$) and \emph{testing} ($30 \%$) sets. Parameter tuning for each of the classification models is performed via $10$-fold cross-validation (within the data from the \emph{training} set). 

The detection (classification) performance was mainly evaluated via the \ac{ROC} \ac{AUC} score. 
However, failure investigation activities associated with a false alarm represent a considerable operational cost for telco providers. Consequently, we also computed the \acp{FPR} as a performance index.

In Fig. \ref{figure:process}, we explain the ML framework for failure detection. Notice that gray boxes represent processes or actions and that white boxes represent their products, intermediate products or inputs. The process starts with the \ac{RSS} measurements, by probing in a deployed network or by simulation. In our implementation, \ac{RSS} values were computed for each antenna in each location $g \in \mathcal{G}$ and used to construct the priority lists $\mathbf{s}_g$. After the lists were created, each pair of \acp{BS} was considered one at a time, and the list of each location was checked to see if it contained the pair of \acp{BS} and in which positions of priority. By observing these positions, a practitioner could compute each of the $u-v$ proximities for the pair of \acp{BS}. Once the practitioner decided which neighborhood category to consider, $u-v$ proximities of each pair of \acp{BS} can be used to determine whether or not they are neighbors. 
When the feature vector was built for a specific target \ac{BS}, the \acp{KPI} vectors of its neighbors were concatenated to the \acp{KPI} vector of the target BS. The process was repeated for all the \acp{BS}, for all the time intervals under study,  to build the data set. Then, the training process took place by minimizing some norm of the difference between the prediction and the real target value. The target value was obtained from the simulation parameters. If there was an ongoing failure in the BS in a specific interval, the target value was defined as 1. Otherwise, the target value was 0.

\subsection{Implementation and scalability details}\label{subsection:implementation}

\begin{figure*}[h!] 
	\centering
	\includegraphics[width=6in]{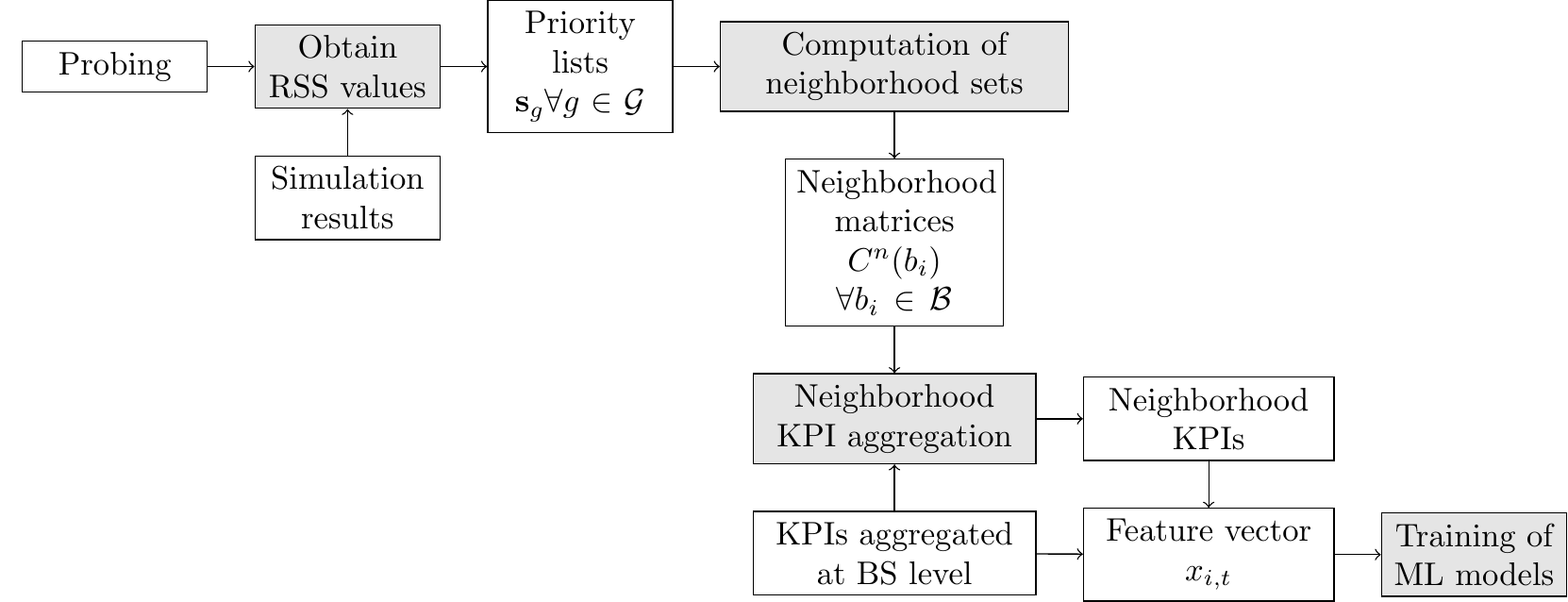}
	\centering	
	\caption{Implementation process.}
	\label{figure:implementation}
\end{figure*}

The proposed framework is based on four main processes (see Fig. \ref{figure:implementation}):

\begin{enumerate}
	\item Retrieval of RSS values at each potential device location and creation of priority lists.
	\item Computation of neighborhood sets based on the priority lists.
	\item Data aggregation (both at the \ac{BS} level and of neighborhoods based on the neighborhood matrices).
	\item Training and evaluation of \ac{ML} models using the aggregated data.
\end{enumerate}

The real-world application of the framework requires some efforts from an operator:
\begin{enumerate}
	\item \ac{RSS} retrieval is a process that can be done via probing a deployed network at each of the potential IoT device locations or via simulation of the \acp{BS}' signal propagation.
	\item Neighborhood computation requires finding the $u-v$ proximities for each pair of \acp{BS} and the further creation of neighborhood sets for each neighborhood category.
	\item Data aggregation. 
	\item All the \acp{ML} models to be trained and evaluated are well-known models, whose complexities and difficulties are well known, and there exists a plethora of pipe-line design strategies that can be used.
\end{enumerate}

In what follows, the first two operator's challenges are analyzed.

To obtain the \ac{RSS} measurements, the telco operator would first need to identify the potential \ac{IoT} device locations and undertake a project to probe, at each location, the signal strength of the \acp{BS} in the range of each location. 
An alternative approach would be to use simulations, as in this work. 
Instead of considering only the potential locations for devices, we divided the space of the city into a grid of squares. 
The positions of the \ac{IoT} devices inside a grid square are approximated with the center of the grid square. 
The \ac{RSS} of every antenna of every \ac{BS} in the simulated network was computed taking into account the antenna tilt, orientation and power, as well as distances and frequencies.

The outcome of this simulation is a data structure of length $\xi G$ (worst-case scenario), where $G$ is the total number of squares in the grid and $\xi$ is the size of the priority lists of antennas at each square. Each item in the structure is a vector containing information regarding the location of the square, the identification of both the \ac{BS} and the antenna and the \ac{RSS} value. 

Another important implementation task is the computation of the neighborhood matrices. To study the scalability of this process so that an operator can estimate its feasibility, we show the computational complexity associated with the process. 
Assuming that $M$ is the total number of \acp{BS} in the city, the computational complexity of the procedure is given by the following polynomial expression:

\begin{equation}
\label{equation:complexity_neighborhood}
	O(M^2+G)
\end{equation}

A distance-based approach, in contrast, involves the computation of the Voronoi regions around each of the \acp{BS}, a process the computational complexity of which is, in general \cite{aurenhammer91}:

\begin{equation}
\label{equation:complexity_voronoi}
	O(M\log{M})
\end{equation}

Compared to Equation (\ref{equation:complexity_voronoi}), the complexity of the neighborhood matrices can be considered more computationally expensive. However,  being polynomial, the method can be considered scalable. If the number $M$ of \acp{BS} of the network is considered constant, Equation (\ref{equation:complexity_neighborhood}) becomes $O(G)$, which is linear with respect to the number $G$ of potential IoT device locations. 

\section{Numerical results}\label{section:results}

\begin{table}[h]
\centering
\caption{Minimum and average \ac{ROC} \ac{AUC} for each classifier.}

\begin{tabular}{lcc}
                 
                                                   & \multicolumn{2}{c}{ROC AUC}                                                   \\
\multicolumn{1}{l|}{\textbf{Classification model}} & \multicolumn{1}{l}{\textbf{Average}} & \multicolumn{1}{l}{\textbf{Minimum}}   \\ \cline{1-3}
\multicolumn{1}{l|}{\emph{Extra Trees}}           & \textbf{0.997}   & 0.971 \\
\multicolumn{1}{l|}{\emph{Random Forests}}        & 0.985            & 0.921 \\
\multicolumn{1}{l|}{\emph{Decision Trees}}        & 0.984            & 0.897 \\
\multicolumn{1}{l|}{\emph{Naive Bayes}}           & 0.983            & 0.938 \\
\multicolumn{1}{l|}{\emph{Bagged Decision Trees}} & 0.981            & 0.879 \\
\multicolumn{1}{l|}{Linear \ac{SVM}}       & 0.959            & 0.912 \\
\multicolumn{1}{l|}{Shallow Neural Network}& 0.955            & 0.898 \\
\multicolumn{1}{l|}{Quadratic \ac{SVM}}    & 0.954            & 0.891 \\
\multicolumn{1}{l|}{Logistic Regression}   & 0.954            & 0.877 \\
\multicolumn{1}{l|}{\ac{RBF} \ac{SVM}}     & 0.953            & 0.902 \\
\multicolumn{1}{l|}{Cubic \ac{SVM}}        & 0.952            & 0.875                                 
\end{tabular}
\label{table:auc_per_model}
\end{table}

We now discuss the results obtained after using four classifiers in the following task: to determine if a \emph{specific} vector of aggregated \acp{KPI} taken during a time interval at a specific BS was produced \emph{or not} during an ongoing failure. This was achieved without knowledge of the past behavior of the BS. 

According to their classification performance both from the point of view of the \ac{ROC} \ac{AUC} and \ac{FPR}, in all of our experiments, we could identify two groups of supervised classifiers:

\begin{itemize}
	\item \emph{Group 1}: consisting of all the \ac{SVM} classifiers, along with Logistic Regressions and the Shallow Neural Networks, which achieve on average an \ac{AUC} score below $0.981$.
	\item \emph{Group 2}: consisting of the ensemble learners (Bagged Decision Trees, Random Forests and Extra Trees), Decision Trees and Naive Bayes, the average \ac{AUC} of which is higher than $0.98$. The Extra Trees classifiers in particular, in the worst performance, achieved an \ac{AUC} higher than $0.97$, and the average score was above $0.99$. We indicate the classifiers of this group with \emph{italics} in Table \ref{table:auc_per_model}.
\end{itemize}

\begin{figure}[!t] 
	\centering
	\captionsetup{justification=centering}
	\includegraphics[width=2.7in]{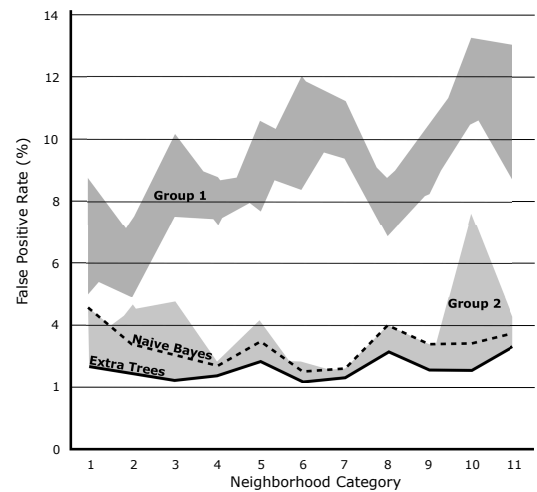}
\centering	
	\caption{Effect of neighborhood category on false positive rate per classifier.}
	\label{fig:proximity_effect_fpr}
\end{figure}

The behavior of the performance of these groups is shown in Figs. \ref{fig:proximity_effect_fpr}, \ref{fig:aggregation_effect}, and \ref{fig:proximity_effect}. It can be argued that the pattern is easily separable, as a low-complexity classifier like Naive Bayes performs very well. While Extra Trees is not a simple classifier, its building strategy is less prone to overfitting than traditional one-hidden-layer Neural Networks and kernel-based methods (\acp{SVM}), which might explain its dominance over those models.

It is important to keep in mind that these \ac{AUC} and \ac{FPR} values are obtained without \emph{any} \ac{BS}-related information (\ac{BS} ID, time, coordinates, number of devices producing traffic, application types, etc).

Among all the experiments, the average \emph{effect of increasing the traffic intensity is mild} (never higher than $1 \%$), though in most classifiers, the effect is slightly negative. Extra Trees is an exception, showing a slightly positive average reaction.

\subsection{Effect of neighborhood category}\label{subsection:results_category} 

In Fig.  \ref{fig:proximity_effect_fpr} we show the \ac{FPR} observed when applying the classification models on the 11 data sets generated by aggregating the simulated data considering the 11 neighborhood matrices (for categories $1,...,11$). The reader should note that the neighborhood definition affects which (and how many) \acp{BS}' \acp{KPI} are included to detect the failure. 
The \ac{FPR} computed for each classifier in this graph is the average of the \acp{FPR} obtained under the two traffic levels for each of the time aggregation sizes, to observe only the effect of the change in the neighborhood definition. 

It is notable that, in general, increasing the category of the neighborhood definition has a detrimental effect on the \ac{FPR} values for classifiers of group 1. The \ac{FPR} values for classifiers of group 2, on the other hand, exhibit neither a clear improvement nor a clear deterioration when increasing the size of the neighborhoods considered.
The results suggest that there is no justification for using neighborhoods of categories higher than 3 when considering \acp{FPR} values.

\begin{figure}[!t] 
	\centering
	\captionsetup{justification=centering}
\includegraphics[width=2.7in]{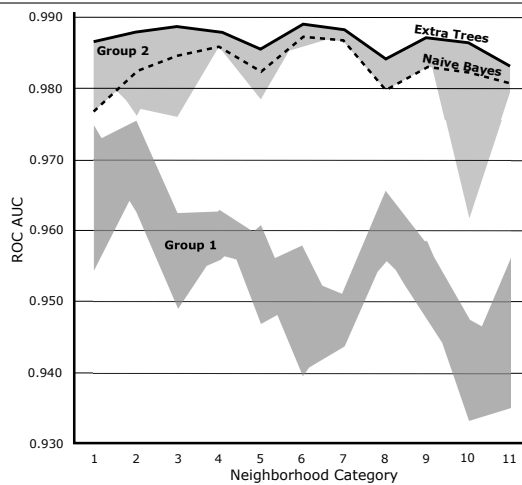}
	\centering	
	\caption{Effect of neighborhood category on \ac{AUC} for each classifier.}
	\label{fig:proximity_effect}
\end{figure}

In Fig. \ref{fig:proximity_effect}, we observe the response of the \ac{ROC} \ac{AUC} values with respect to changes in the neighborhood definition. A clear distinction in the behavior of classifiers of Groups 1 and 2 can also be observed in terms of the \ac{AUC}, and group 2 does not appear to benefit from having a neighborhood category higher than 3. The performance of classification models in group 1 also deteriorates in terms of the \ac{AUC}, progressively losing the separation ability as the neighborhood size increases.

\subsection{Effect of aggregation size}\label{subsection:results_aggregation} 

\begin{figure}[!t] 
	\centering
	\captionsetup{justification=centering}
	\includegraphics[width=2.7in]{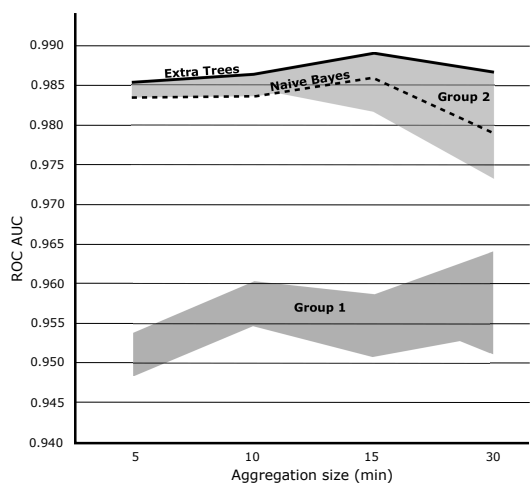}
\centering	
	\caption{Effect of aggregation on \ac{AUC} for each classifier.}
	\label{fig:aggregation_effect}
\end{figure}

In Fig. \ref{fig:aggregation_effect}, we show how the average \ac{ROC} \ac{AUC} values for each group of classifiers responds to increasing sizes of the aggregation time bins. To build this figure, we averaged the \ac{AUC} values obtained by each classifier over the data generated under the two levels of traffic and preprocessed under the 11 neighborhood categories to observe only  the effect produced by the size of the aggregation bins.

We find that increasing aggregation size from 5 to 10 minutes has effects that range from mild (for classifiers of Group 2, which already have \ac{AUC} values higher than $0.97$) to clearly positive (for classifiers of Group 1) (see Fig. \ref{fig:aggregation_effect}). With the exception of Bagged Trees, all models in Group 2 clearly benefit from the increase in aggregation from 10 to 15 minutes. Increasing the aggregation to 30 minutes, however, appears to blur the patterns and provoke a deterioration of their performance. 
Models of Group 1 had no performance improvement when augmenting the aggregation size to 15 minutes and had no uniform response to the increase in aggregation size to 30 minutes. 

When averaging to observe the \ac{ROC} \ac{AUC} and \ac{FPR} responses to all the neighborhood categories and all the aggregations sizes, 
Extra Trees consistently showed the best performance. Naive Bayes, being a less complex classifier, had similar results on average, and might be a sound enough choice for an operator in the scenarios at hand.

\subsection{Interaction between aggregation size and neighborhood category}\label{subsection:results_interaction}

\begin{figure}[!t] 
	\centering
	\captionsetup{justification=centering}
	\includegraphics[width=2.7in]{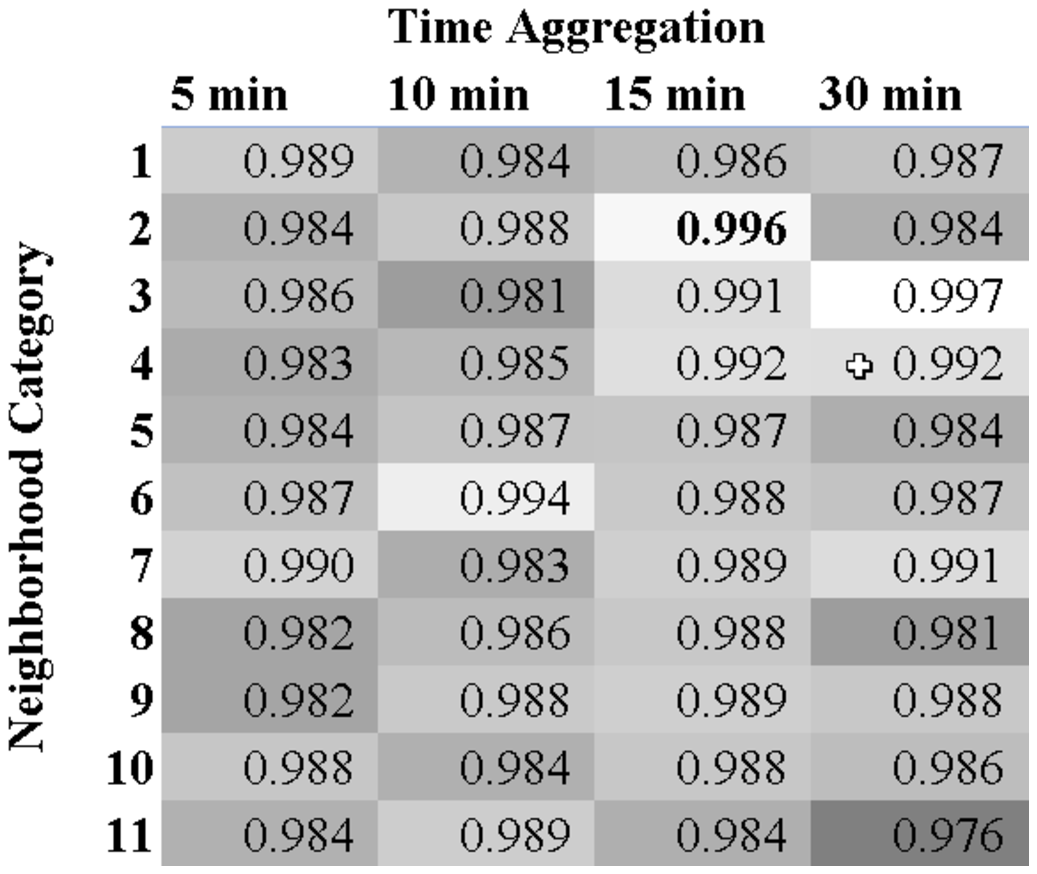}
	\centering	
	\caption{Joint effect of proximity and aggregation levels on Extra Trees \ac{AUC} score.}
	\label{fig:agg_vs_prox_xtrees}
\end{figure}

\begin{figure}[!t] 
	\centering
	\captionsetup{justification=centering}
	\includegraphics[width=2.7in]{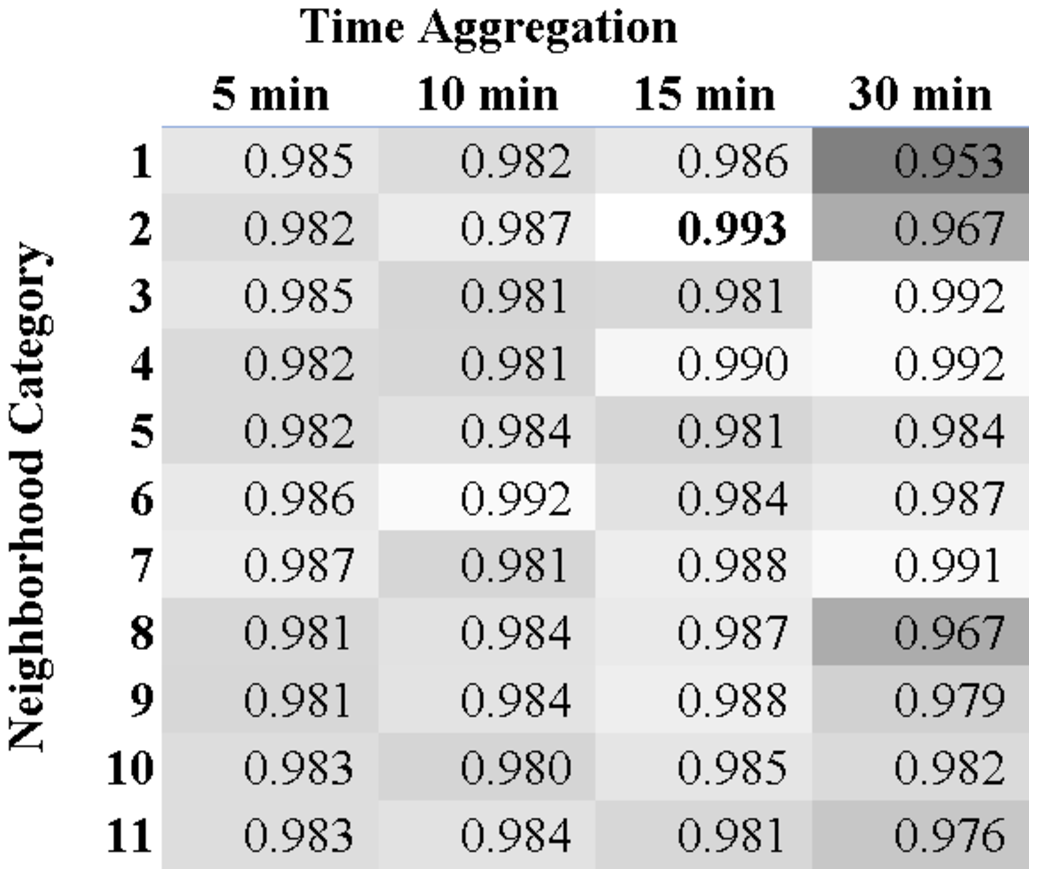}
	\centering	
	\caption{Joint effect of proximity and aggregation levels on Naive Bayes \ac{AUC} score.}
	\label{fig:agg_vs_prox_nbayes}
\end{figure}

Figs. \ref{fig:agg_vs_prox_xtrees} and \ref{fig:agg_vs_prox_nbayes} show the average \ac{AUC} scores for the two best models: Extra Trees and Naive Bayes, respectively. In these figures, to analyze the \emph{joint effect of time aggregation size and neighborhood category},  we created a heatmap, in which lighter colors represent higher \ac{AUC} scores and consequently better detection performance.

To compute these values, the \ac{AUC} values were averaged among only the two traffic levels to observe the interaction between the neighborhood definition and time aggregation. 

In the aforementioned figures, it can be noted that both methods produced \ac{AUC} scores close to 1, making evident that the models have a high separation capacity for these data sets. However, neighborhood categories 2, 3 and 4 obtained the \emph{best} scores, especially when the size of the time aggregations was 10, 15 and 30. 

In particular, the best results were achieved with 15 minutes of aggregation and neighborhood category 2, allowing the Extra Trees classifiers to achieve an \ac{AUC} score of $0.996$, and the Naive Bayes classifier a score of $0.993$.

\section{Conclusions and future work}\label{section:conclusion}

In this paper, we proposed a supervised learning framework to detect \ac{RACH}-related \emph{sleeping} cells in a smart-city cellular infrastructure. We used well-known binary classification techniques to detect network elements at fault based on the analysis of aggregated \acp{KPI} such as the \ac{RACH} collision probability and the delay.

\ac{RACH}-related sleeping cells are difficult to detect due to the lack of evidence in the \acp{KPI} from a faulty cell. 
To overcome this problem, we have proposed to jointly consider the \acp{KPI} of one cell with those from the neighboring cells. 
We have also proposed a novel definition for neighbors of a cell, not choosing the nodes geographically closer to a cell but rather those that would be more likely impacted by its failure.

We used data obtained from a large-scale IoT network simulator that employs real data on the telecommunication infrastructure and on the position of IoT nodes in a smart-city environment. Although \ac{LTE} was chosen to obtain numerical results, the proposed framework can easily be adapted to other cellular technologies, such as 5G.

Different time aggregation interval sizes were tested for the \acp{KPI}: 15 minutes resulted in the aggregation interval that permitted achieving the highest \ac{AUC}. This aggregation level permits heavily reducing the amount of data to be analyzed by a network operator to detect faulty elements, resulting in large potential savings. The numerical results also proved Extra Trees and Naive Bayes to be the most effective binary classification techniques among the ones considered in this work. Broadly speaking, the results suggest that simple and ensemble models, known for being less prone to overfitting, are superior to kernel models and neural networks.

The preprocessing approach based on aggregations and the inclusion of information regarding the ``neighborhood'' of a \ac{BS} has shown its value in the classification task. We are currently working on how to adapt this strategy for forecasting and anomaly detection.

\acrodef{ARMA}{Auto-regressive moving average}
\acrodef{ARIMA}{Auto-regressive integrated moving average}
\acrodef{DMC}{Data Management Center}
\acrodef{EIRP}{Equivalent Isotropic Radiated Power}
\acrodef{HTC}{Human Type Communication}
\acrodef{KPI}{Key Performance Indicator}
\acrodef{IoT}{Internet of Things}
\acrodef{M2M}{Machine-to-Machine}
\acrodef{MTC}{Machine Type Communication}
\acrodef{BS}{Base Station}
\acrodef{CQI}{Channel Quality Indicator}
\acrodef{SON}{Self-Organizing Network}
\acrodef{UE}{User Equipment} 
\acrodef{OSS}{Operation Support System}
\acrodef{RACH}{Random-Access Channel}
\acrodef{RAO}{Random-Access Opportunity}
\acrodef{RB}{Resource Block}
\acrodef{RSRP}{Reference Signals Received Power}
\acrodef{RSRQ}{Reference Signals Received Quality}
\acrodef{SOM}{Self Organizing Map}
\acrodef{SMOTE}{Synthetic Minority Over-Sampling Technique}
\acrodef{SVM}{Support Vector Machine}
\acrodef{QoS}{Quality of Service}
\acrodef{QoE}{Quality of Experience}
\acrodef{microPMU}{Micro-Phasor Measuring Unit}
\acrodef{L-SVM}{Linear Support Vector Machine}
\acrodef{BDT}{Bagged Decision Trees}
\acrodef{PMU}{Phasor Measurement Unit}
\acrodef{IoT}{Internet of Things}
\acrodef{HTC}{Human-Type Communications}
\acrodef{AUC}{Area Under the Curve}
\acrodef{ROC}{Receiver Operating Characteristic}
\acrodef{BS}{Base Station}
\acrodef{FPR}{False Positive Rate}
\acrodef{RBF}{Radial Basis Function}
\acrodef{RSS}{Received Signal Strength}
\acrodef{LTE}{Long-Term Evolution}
\acrodef{3GPP}{$3^{rd}$ Generation Partnership Project}
\acrodef{4G}{$4^{th}$ Generation of broadband cellular network technology}
\acrodef{5G}{$5^{th}$ Generation}
\acrodef{ML}{Machine Learning}
\acrodef{EPC}{Evolved Packet Core}

%
%
%
%
%

\bibliographystyle{IEEEtran}

\bibliography{new_telecom,new_toto}

\begin{IEEEbiography}[{\includegraphics[width=1in,height=1.25in,clip,keepaspectratio]{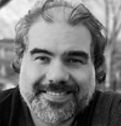}}]{Orestes G. Manzanilla-Salazar} (M'19) received his B.S. (2002) in production engineering and his M.S. (2005) in systems engineering from Simón Bolívar University (USB) in Venezuela. He is currently working on his Ph.D in electrical engineering at Ecole Polytechnique de Montreal, Montreal, QC, Canada. 

From 2008 to 2015, he was an Assistant Lecturer at USB. He taught courses in the areas of mathematical programming, decision science and operations management and focused his research on linear and integer programming models to create supervised classifiers. He worked 6 years in several positions for private firms in the areas of engineering, technology and finance. His present research focus is on machine learning and analytics tools for telco networks management and on network-enabled systems.
\end{IEEEbiography}

\begin{IEEEbiography}[{\includegraphics[width=1in,height=1.25in,clip,keepaspectratio]{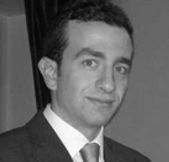}}]{Filippo Malandra} (M'17) received B.Eng. and M.Eng. degrees in telecommunications engineering from the Politecnico di Milano, Milan, in 2008 and 2011, respectively, and a Ph.D. degree in electrical engineering from Ecole Polytechnique de Montreal, Montreal, QC, Canada, in 2016. 

He is currently an Assistant Professor of Research with the Department of Electrical Engineering at the University at Buffalo. His research interests include the performance analysis and simulation of mobile networks (4G/5G) for smart grids, smart cities and IoT applications.
\end{IEEEbiography}

\begin{IEEEbiography}[{\includegraphics[width=1in,height=1.25in,clip,keepaspectratio]{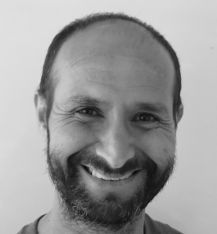}}]{Hakim Mellah} received his B.S (1996) and M.Sc.(1999) in electrical engineering and his Ph.D. in electrical and computer engineering from Concordia University, Montreal, QC, Canada, in 2018. 

He is currently a postdoc at Ecole Polytechnique de Montreal. His research interests focus mainly on machine quality of experience for IoT applications in a smart-city environment. 
\end{IEEEbiography}

\begin{IEEEbiography}[{\includegraphics[width=1in,height=1.25in,clip,keepaspectratio]{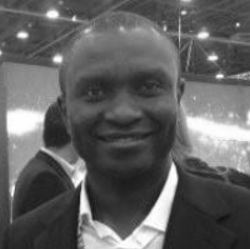}}]{Constant Wette} holds an M.Sc. in computer engineering from Ecole Polytechnique de Montreal, Montreal, QC, Canada (2000), an M.Eng. in electrical engineering from Ecole Polytechnique de Yaoundé, Cameroon (1996), an MBA from HEC Montreal, Montreal, QC, Canada (2010) and a Graduate Certificate in data science from Harvard University (2016). 

He is a Senior System Developer at Business Area Digital Services, Ericsson. He joined Ericsson in 2000 and has led research projects, product development, innovation and new business development in different technology domains including data analytics, IoT/M2M, IMS and telecommunication cloud architectures.
\end{IEEEbiography}

\begin{IEEEbiography}[{\includegraphics[width=1in,height=1.25in,clip,keepaspectratio]{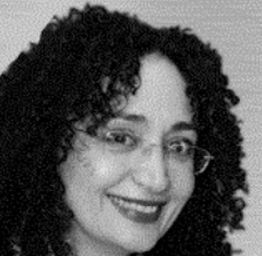}}]{Brunilde~Sans\`o} (M’92–SM’17) is a Full Professor of telecommunication networks with the Department of Electrical Engineering, Ecole Polytechnique de Montreal, and the Director of the LORLAB, a research group dedicated to developing methods for the design and performance of wireless and wireline telecommunication networks. 

Over her career of more than 25 years, she has received several awards and honors, has published extensively in the telecommunication networks and operations research literature, and has been a consultant for telecommunication operators, equipment and software manufacturers, and the mainstream media.
\end{IEEEbiography}

\EOD

\end{document}